\documentclass[twocolumn,superscriptaddress]{revtex4}

\usepackage{amsmath}
\usepackage[final]{graphicx}

\begin{document}

\title{Charge transfer between epitaxial graphene and silicon carbide}
\author{Sergey Kopylov}
\affiliation{Department of Physics, Lancaster University, LA1 4YB Lancaster, UK}

\author{Alexander Tzalenchuk}
\affiliation{National Physical Laboratory, TW11 0LW Teddington, UK}

\author{Sergey Kubatkin}
\affiliation{Department of Microtechnology and Nanoscience, 
Chalmers University of Technology, S-412 96 Goteborg, Sweden}

\author{Vladimir I. Fal'ko}
\affiliation{Department of Physics, Lancaster University, LA1 4YB Lancaster, UK}

\date{\today}

\begin{abstract}
We analyse doping of graphene grown on SiC in two models which differ by the
source of charge transfered to graphene, namely, from SiC surface and from
bulk donors. For each of the two models, we find the maximum electron
density induced in monolayer and bilayer graphene, which is determined by
the difference between the work function for electrons in pristine graphene
and donor states on/in SiC, and analyse the responsivity of graphene to the
density variation by means of electrostatic gates.
\end{abstract}

\maketitle

Graphene \cite{Geim} - monolayer or bilayer of carbon atoms with a honeycomb
lattice - is a gapless semiconductor, which can be used as a
current-carrying element in field-effect transistors. It has been found that
the epitaxial graphene grown onto cm-size wafers of the Si-terminated face
of SiC \cite{Ohta,Yakimova,Emtsev} maintains structural integrity over
large area and demonstrates a relatively high mobility of carriers
\cite{Tza,First,Lin}. This makes graphene synthesized on SiC (SiC/G) a promising
platform to build integrated electronic circuits, assuming one can control
the carrier density in it. For transistor applications, bilayer
graphene in SiC/G is a particularly interesting material, since interlayer
asymmetry (\textit{e.g.}, induced by a transverse electric field) opens a
minigap in its spectrum \cite{BilayerGrapheneSpectrum,BLG-exp}.

In this Letter we present a theory of charge transfer from SiC to the
epitaxial monolayer (MLG) and bilayer (BLG) graphene grown on its surface.
It has been noticed that epitaxial graphene is
always substantially n-doped, so that use of SiC/G in transistors requires
reduction of carrier density using gates \cite{Emtsev}.
The initial doping of graphene comes from a combination of bulk donors in SiC
with the volume density $\rho$ and surface donor states with the sheet density of
surface states, $\gamma $ (DoS). The charge transfer to graphene in a
top-gated field-effect transistor can be found from solving two coupled
equations: 
\begin{gather}
\gamma [ A-4\pi e^{2}d(n+n_{g})-\varepsilon_{F}(n)]+\rho l=n+n_{g},
\label{FullSet1} \\
\tilde{A}=\varepsilon _{F}(n)+U+4\pi e^{2}d(n+n_{g}).  \label{FullSet2}
\end{gather}
Equation (\ref{FullSet1}) accounts for the charge balance, with $%
n_{g}=CV_{g}/e$ ($e>0$) being the areal density of electrons transferred
to the gate. Here, $\tilde{A}$ /$A$ is the difference between the
work function of graphene and the work function of electrons in the bulk/surface donors in
SiC, and $\varepsilon_{F}$ is the Fermi energy in doped graphene. Equation
(\ref{FullSet2}) describes the equilibrium between electrons in
graphene and bulk donors, with $l$ standing for the depletion layer width in
SiC, $U=\frac{2\pi e^{2}}{\chi }\rho l^{2}$ being the height of the Shottky
barrier ($\chi$ is dielectric constant of SiC), and $d$ - the
distance between the SiC surface and the middle of graphene layer.

In the following, we calculate the density $n$ of electrons (for both MLG
and BLG) in two limits: graphene doping dominated by the charge transfer
from \textbf{(a) surface donors}, which corresponds to solving Eq. (\ref{FullSet1})
with $\rho l\longrightarrow 0$, and \textbf{(b) bulk donors} (e.g. nitrogen),
which corresponds to solving Eqs. (\ref{FullSet1},\ref{FullSet2}) with
$\gamma =0$. Charge transfer in a more generic situation, with arbitrary
$\rho $ and $\gamma $, can be assessed by taking the largest of the two
estimates. Then, we determine the responsivity factor, $r=-dn/dn_{g}$,
which characterizes the ability to control the carrier density using
external gates: $r\longrightarrow 1$ would be optimal for transistor
operation of gated SiC/G devices whereas 
$r\ll 1$ would indicate that transistor operation is impossible.

\begin{figure}[tbp]
\centering \includegraphics[width=.53\columnwidth]{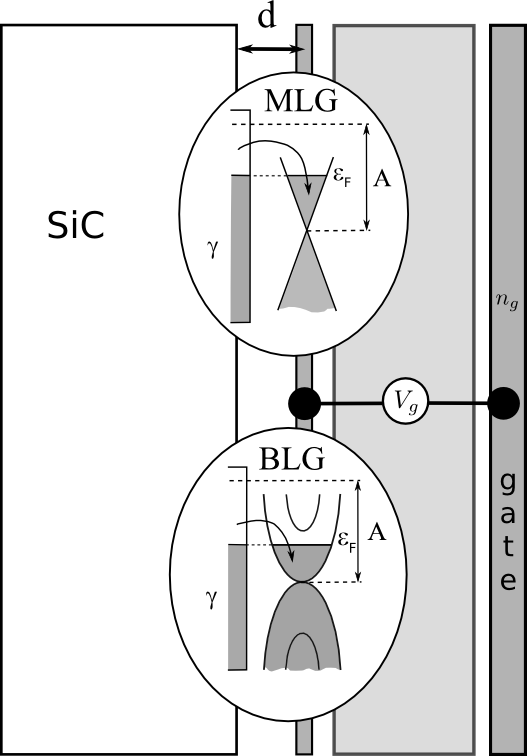}
\caption{Sketch of a SiC/G-based field effect transistor. Insets illustrate
relevant part of the MLG and BLG band structure.}
\label{fig:setup}
\end{figure}

\textbf{Monolayer graphene} (MLG) has linear spectrum
$\varepsilon _{\pm}(p)=\pm vp$, in the two valleys, corresponding to the non-equivalent
corners $K$ and $K^{\prime }$ of the hexagonal Brillouin zone, so that
$\varepsilon _{F}(n)=\hbar v\sqrt{\pi n}$ (we take into account both
valley and spin degeneracy of the electron states). In the limit (a)
we find that the carrier density is 
\begin{gather}
n^{I}=\frac{A\gamma ^{2}\left[ \sqrt{1+\frac{\gamma _{A}}{\gamma
^{2}}\left( \gamma -\frac{n_{g}}{A}\frac{\gamma _{d}+\gamma }{\gamma _{d}}%
\right) \frac{\gamma _{d}+\gamma }{\gamma _{d}}}-1\right] ^{2}}{\gamma
_{A}\left( 1+\frac{\gamma }{\gamma _{d}}\right) ^{2}}, \nonumber \\
\gamma _{d}=\frac{1}{4\pi de^{2}},\qquad \gamma _{A}=\frac{4A}{\pi \hbar^{2}v^{2}}. \label{dens-mlg}
\end{gather}
The initial density of electrons in graphene is described by Eq.~(\ref{dens-mlg})
($n_{g}=0$) with two characteristic regimes,
\begin{equation}
n^{I}\approx \left\{
\begin{array}{ll}
A\gamma , & \gamma \ll \gamma _{\ast }^{I}; \\ 
n_{\ast }^{I}, & \gamma \gg \gamma _{\ast }^{I},
\end{array}
\right.
\label{nIsurf}
\end{equation}
where $n_{\ast }^{I}$ and $\gamma_{\ast}^{I}$ are, respectively, the
saturation value for the carrier density in SiC/G and the crossover value of
DoS of donors on the SiC surface at which $n(\gamma)$ saturates:

\begin{equation}
\gamma _{\ast }^{I}=\frac{\gamma _{A}}{\left( 1+\sqrt{1+\frac{\gamma _{A}}{
\gamma _{d}}}\right) ^{2}},\quad n_{\ast }^{I}=\frac{A\gamma _{A}}{\left(
1+\sqrt{1+\frac{\gamma _{A}}{\gamma _{d}}}\right) ^{2}}.  \label{sat-mlg}
\end{equation}

The responsivity of MLG in SiC/G is, then, 
\begin{equation}
\begin{array}{c}
r|_{n_{g}=0}=1-\left( 1+\frac{\gamma _{A}}{\gamma _{d}}+\frac{\gamma _{A}}{
\gamma }\right) ^{-1/2}\approx \\ 
\approx \left\{ 
\begin{array}{cc}
1, & \gamma \ll \gamma _{\ast }^{I}; \\ 
1-\frac{1}{\sqrt{1+\gamma _{A}/\gamma _{d}},} & \gamma \gg \gamma _{\ast}^{I}.
\end{array}
\right.
\end{array}
\label{susc-mlg}
\end{equation}

\begin{figure}[tbp]
\centering \includegraphics[width=.85\columnwidth]{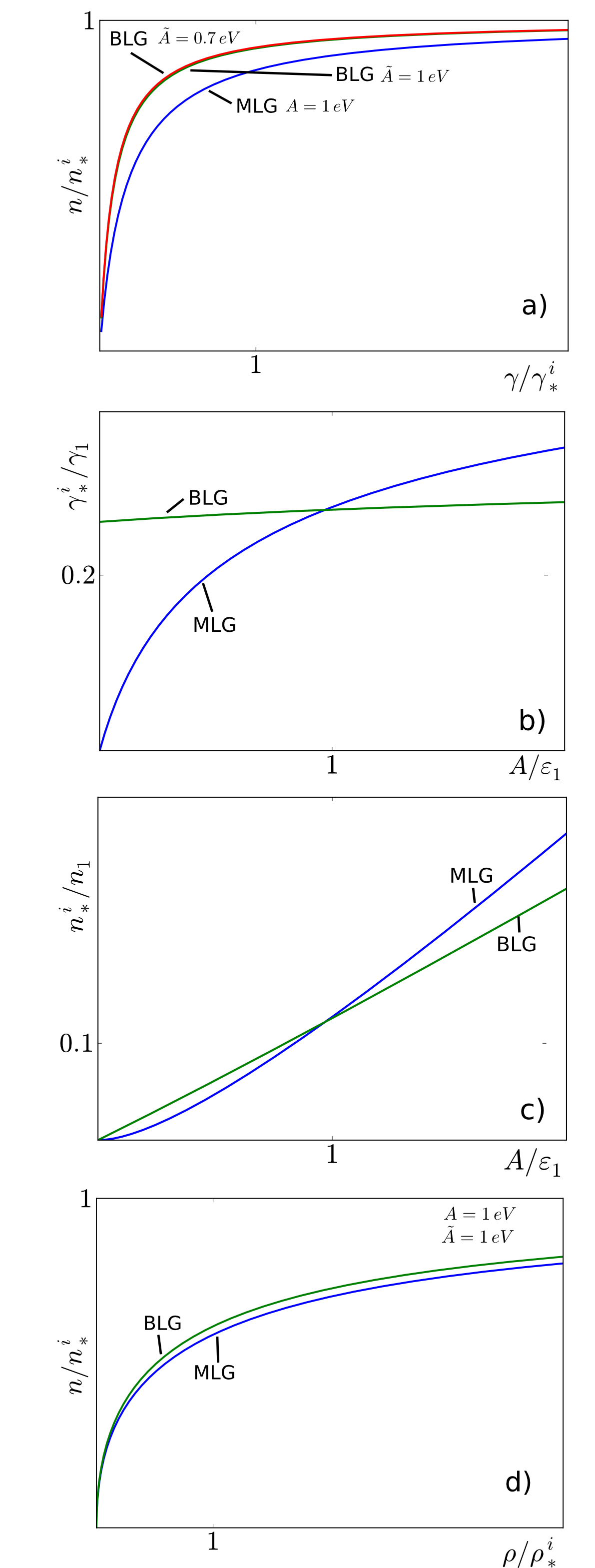}
\caption{Comparison between charge transfer from SiC to MLG and BLG, with $A$
measured in units of split-band energy in BLG, $\varepsilon_{1}=0.4$~eV,
$d=0.3$~nm for MLG and $d=0.5$~nm for BLG. (a) Electron
concentration in graphene $n$ dominated by charge transfer from surface states. (b)
Values $\gamma_{\ast}$ of the surface DoS at which $n(\gamma)$
saturates. (c) Saturation density value as a function of $n_{\ast}$,
in units of $n_{1}$. (d) Electron bulk donors density $\rho$ for $\gamma=0$.}
\label{fig:plots}
\end{figure}

In the limit (b) we find $n(\rho )$ by solving Eqs.~(\ref{FullSet1},\ref{FullSet2})
numerically. The numerical solution shown in Fig.~\ref{fig:plots}
(d) interpolates between the regimes of weak and strong graphene doping: 
\begin{gather}
n^{I}\approx \left\{ 
\begin{array}{ll}
\sqrt{\frac{\tilde{A}\chi \rho }{2\pi e^{2}}}, & \rho \ll \rho _{\ast }^{I};
\\
\tilde{n}_{\ast }^{I}, & \rho \gg \rho _{\ast }^{I};
\end{array}
\right.   \label{n_blk_mlg} \\
\tilde{n}_{\ast }^{I}=\frac{\tilde{A}\tilde{\gamma}_{A}}{\left( 1+\sqrt{1+
\frac{\tilde{\gamma}_{A}}{\gamma _{d}}}\right) ^{2}},\;\tilde{\gamma}_{A}=
\frac{4\tilde{A}}{\pi \hbar ^{2}v^{2}},  \notag \\
\rho _{\ast }^{I}=\frac{2\pi e^{2}\tilde{A}\tilde{\gamma}_{A}^{2}}{\chi
\left( 1+\sqrt{1+\frac{\tilde{\gamma}_{A}}{\gamma _{d}}}\right) ^{4}}.
\notag
\end{gather}%
The responsivity, $r|_{n_{g}=0}$ of MLG in SiC/G in the limit (b) can be
described using Eq. (\ref{susc-mlg}), but with $\gamma _{A}$ replaced by
$\tilde{\gamma}_{A}$ and the upper/lower limits corresponding to $\rho \ll
\rho _{\ast }^{I}$ and $\rho \gg \rho _{\ast }^{I}$, respectively.

\textbf{Bilayer graphene} (BLG) has the spectrum~\cite{BilayerGrapheneSpectrum}
with conduction (+) and valence (-) bands, which,
in the vicinity of the Brillouin zone corners, are described by 
\begin{equation*}
\varepsilon _{s\pm }(p)=\pm \left( \sqrt{\frac{\varepsilon _{1}^{2}}{4}
+v^{2}p^{2}}+\frac{s\varepsilon _{1}}{2}\right) ,
\end{equation*}
where, $\varepsilon _{1}$ is the interlayer coupling and $s$ distinguishes
between pairs of degenerate ($s=-1$) and split ($s=1$) bands. The Fermi
level in n-doped BLG with sheet electron density $n$ is determined as 
\begin{equation*}
\varepsilon _{F}=\left\{ 
\begin{array}{ll}
\sqrt{\frac{\varepsilon _{1}^{2}}{4}+\pi \hbar ^{2}v^{2}n}-\frac{\varepsilon_{1}}{2}, & n<n_{1}, \\ 
\hbar v\sqrt{\frac{\pi n}{2}}, & n>n_{1},
\end{array}
\right.
\end{equation*}
where $n_{1}=2\varepsilon _{1}^{2}/(\pi \hbar ^{2}v^{2})\approx 3\cdot 10^{13}$~cm$^{-2}$.

In the limit (a), this gives 
\begin{equation*}
n^{II}=\frac{2A(A+\varepsilon _{1})/(\pi \hbar ^{2}v^{2})}{1+\frac{\tilde{\gamma}_{1}}
{\gamma _{d}}+\frac{\tilde{\gamma}_{1}}{\gamma }+\sqrt{\left( 1+
\frac{\gamma _{1}}{\gamma _{d}}+\frac{\gamma _{1}}{\gamma }\right) ^{2}+
\frac{\gamma _{A}}{\gamma _{d}}+\frac{\gamma _{A}}{\gamma }}}
\end{equation*}
for $n^{II}(\gamma )<n_{1}$ and $n_{g}=0$. Here, $\gamma _{1}=\varepsilon
_{1}/\pi \hbar ^{2}v^{2}$ and $\tilde{\gamma}_{1}=\gamma _{1}+\gamma _{A}/2$.
For larger densities, $n^{II}>n_{1}$, 
\begin{equation*}
n^{II}=\frac{8A^{2}}{\pi \hbar ^{2}v^{2}}\frac{1}{\left( 1+\sqrt{1+
\frac{2\gamma _{A}}{\gamma }+\frac{2\gamma _{A}}{\gamma _{d}}}\right) ^{2}},
\end{equation*}
which resembles Eq.~(\ref{dens-mlg}) for MLG, but with
$\gamma_{A}\rightarrow 2\gamma _{A}$.

Similarly to MLG, the density in BLG on SiC saturates upon the increase of
surface DoS of donors. The crossover to the saturated density, $n_{\ast
}^{II}(A)=A\gamma _{\ast }^{II}(A)$, occurs at 
\begin{equation*}
\gamma _{\ast }^{II}=\left\{
\begin{array}{ll}
\frac{2\gamma _{A}}{\left( 1+\sqrt{1+\frac{2\gamma _{A}}{\gamma _{d}}}
\right) ^{2}}, & \frac{2\gamma _{1}}{\gamma _{d}}+1<\frac{\gamma _{A}}{4\gamma _{1}}; \\ 
\frac{\frac{\gamma _{A}}{2}+2\gamma _{1}}{1+\frac{\gamma _{A}}{2\gamma _{d}}+
\frac{\gamma _{1}}{\gamma _{d}}+\sqrt{\left( 1+\frac{\gamma _{1}}{\gamma _{d}
}\right) ^{2}+\frac{\gamma _{A}}{\gamma _{d}}}}, & \frac{2\gamma _{1}}{
\gamma _{d}}+1>\frac{\gamma _{A}}{4\gamma _{1}}.
\end{array}
\right.
\end{equation*}
The dependence of $n^{II}(\gamma )$, $\gamma _{\ast }^{II}(A)$ and
$n_{\ast}^{II}(A)$ on the relative size of the band splitting $\varepsilon _{1}$,
and the graphene-surface donors work function $A$ is shown in Fig.~\ref{fig:plots} (a-c).
Responsivity of the BLG to the gate voltage is
high or low, depending on whether the saturation regime for the carrier
density is reached, or not. For $\gamma \ll \gamma _{\ast }^{II}$,
$r\approx 1$. For $\gamma \gg \gamma _{\ast }^{II}$, 
\begin{equation}
r\approx \left\{ 
\begin{array}{ll}
1-\frac{1}{\sqrt{1+\frac{2\gamma _{A}}{\gamma _{d}}}}, & \frac{2\gamma _{1}}
{\gamma _{d}}+1<\frac{\gamma _{A}}{4\gamma _{1}}; \\ 
1-\frac{1}{\sqrt{\left( 1+\frac{\gamma _{1}}{\gamma _{d}}\right) ^{2}+
\frac{\gamma _{A}}{\gamma _{d}}}}, & \frac{2\gamma _{1}}{\gamma _{d}}+1>
\frac{\gamma _{A}}{4\gamma _{1}}.
\end{array}
\right.  \label{r-blg}
\end{equation}

In the limit (b), when $n$ is determined by charge transfer from bulk donors in SiC,
\[
n^{II}=\left\{ 
\begin{array}{ll}
\sqrt{\frac{\tilde{A}\chi \rho }{2\pi e^{2}}}, & \rho \ll \rho _{\ast }^{II};
\\ 
\tilde{n}_{\ast }^{II}(\tilde{A}), & \rho \gg \rho _{\ast }^{II};
\end{array}
\right. \quad \rho _{\ast }^{II}=\frac{2\pi e^{2}(\tilde{n}_{\ast }^{II})^{2}
}{\chi \tilde{A}},
\]
with $\tilde{n}_{\ast }^{II}=n_{\ast }^{II}(\tilde{A})$, and responsivity
$r\approx 1$ of BLG requires that $\rho \ll \rho _{\ast }^{II}$, whereas for
$\rho \gg \rho _{\ast }^{II}$ responsivity is described by the same limits as
in Eq. (\ref{r-blg}), with $\gamma _{A}$ replaced by $\tilde{\gamma}_{A}$.

Independently of the number of layers, the gate voltage
$V_{g}^{\ast}$ needed to reach the neutrality point in
graphene controlled by the top gate with mutial capacitance $C$ is
\begin{equation}
V_{g}^{\ast } \approx \frac{e}{C} \, \text{max} \left(
\frac{A}{\gamma^{-1} +\gamma^{-1} _{d}}, 
\frac{2\tilde{A}\gamma _{d}}{1+\sqrt{1+\frac{
8\pi e^{2}\tilde{A}\gamma _{d}^{2}}{\chi \rho }}} \right).
\end{equation}

In summary, we calculated charge transfer from SiC to epitaxial graphene.
In the case when the charge transfer is dominated by donors on the surface
of SiC with $A\sim 1$eV or donors in the bulk of SiC with $\tilde{A}\sim 1$eV,
\cite{Sonde}, we estimate that the saturation density of n-type doping of MLG
is $10^{13}$cm$^{-2}$, which corresponds to $\varepsilon _{F}\approx 0.4$eV
(for $d\approx 0.3$nm). This value of carrier
density occurs when the donor volume density is $\rho >\rho _{\ast }^{I}\sim
10^{19}$~cm$^{-3}$ (we use $\chi \sim 10$ for 6H SiC \cite{Patrick}) or the surface states
have DoS $\gamma >\gamma _{\ast }^{I}\sim 1\cdot 10^{13}$~cm$^{-2}$eV$^{-1}$.
For  lesser doping of SiC, $\gamma<\gamma_*$ and $\rho<\rho_*$,
one should use the larger of the estimates from Eqs.~(\ref{nIsurf},\ref{n_blk_mlg})
This can be compared to the data reported in the recent studies of
epitaxial graphene indicating a substantial initial level of n-type doping
of SiC/G, very often \cite{Emtsev} as high as 10$^{13}$cm$^{-2}$. However,
some particular growth processes produce SiC/G with a much lower doping
level \cite{Weingart,Tza}, indicating that efficient annealing of donors on and near
the SiC surface is possible.

We thank T. Seyller and R. Yakimova for discussions. This work was
supported by EPSRC grant EP/G041954 and EU-FP7 ICT STREP Concept Graphene.


\begin{thebibliography}{99}
\bibitem{Geim} A. K. Geim, Science \textbf{324}, 1530 (2009), arXiv:0906.3799.

\bibitem{Ohta} T. Ohta, \textit{et al}, Science \textbf{313}, 951 (2006); A.
Bostwick, \textit{et al}, Nat. Phys. \textbf{3}, 36 (2007); T. Ohta,
\textit{et al}, Phys. Rev. Lett. \textbf{98}, 206802 (2007).

\bibitem{Yakimova} C. Virojanadara, \textit{et al}, Phys. Rev. B \textbf{78}
, 245403 (2008); G. Guo \textit{et al.}, Appl. Phys. Lett. \textbf{90},
253507 (2007); Y.Q. Wu \textit{et al}., Appl. Phys. Lett. \textbf{92},
092102 (2008); J. Kedzierski \textit{et al}., IEE Trans. Electron. Devices 
\textbf{55}, 2078 (2008).

\bibitem{Emtsev} K. Emtsev, et al, Nature Mater. \textbf{8}, 203 (2009); C.
Coletti, \textit{et al}, Phys. Rev. B \textbf{81} (2010); J. Jobst,
\textit{et al}, Phys. Rev. B \textbf{81} (2010); J.S. Moon, \textit{et al}, IEEE
Electron Device Lett. \textbf{31}, 260-262 (2010).

\bibitem{Tza} A. Tzalenchuk \textit{et al.}, Nature Nanotech. \textbf{5}, 186 (2010), arXiv:0909.1220.

\bibitem{First} P.N. First \textit{et al.}, MRS Bulletin \textbf{35}, 296
(2010).

\bibitem{Lin} Y.M. Lin \textit{et al.}, Nano Letters \textbf{9}, 422 (2009).

\bibitem{BilayerGrapheneSpectrum} E. McCann and V.I. Fal'ko, Phys. Rev.
Lett. \textbf{96}, 086805 (2006), arXiv:cond-mat/0510237.

\bibitem{BLG-exp} J.B. Oostinga \textit{et al.}, Nature Mat. \textbf{7}, 151
(2007); Y. Zhang \textit{et al.}, Nature \textbf{459}, 820 (2009); B.N.
Szafranek, \textit{et al.}, Appl. Phys. Lett. \textbf{96}, 112103 (2010).

\bibitem{Bardin} J. Bardin, Phys. Rev. \textbf{71}, 717 (1947).

\bibitem{Weingart} S. Weingart, \textit{et al}, Appl. Phys. Lett. \textbf{95}
262101 (2009), arXiv:0910.4010.

\bibitem{Sonde} S. Sonde \textit{et al}., Phys. Rev. B \textbf{80}, 241406
(2009).

\bibitem{Patrick} L. Patrick, W. J. Choyke, Phys. Rev. B \textbf{6}, 2255 (1970).

\end{thebibliography}
\end{document}